\documentstyle[preprint,aps]{revtex}
\preprint{USM-TH-149}
\begin{document}
\title{A New Approach to Scale Symmetry Breaking and Confinement}
\author{P. Gaete$^{1}$\thanks{
E-mail: patricio.gaete@fis.utfsm.cl} and E. I. Guendelman
$^{2}$\thanks{ E-mail: guendel@bgumail.bgu.ac.il}}
\address{$^1$Departamento de F\'{\i}sica, Universidad T\'{e}cnica
F. Santa Mar\'{\i}a, Casilla 110-V, Valpara\'{\i}so, Chile. \\
$^2$Physics Department, Ben Gurion University, Beer Sheva 84105,
Israel.} \maketitle

\begin{abstract}
Scale invariant theories which contain (in $4-D$) a four index
field strength are considered. The integration of the equations of
motion of these $4-index$ field strength gives rise to scale
symmetry breaking. The phenomena of mass generation and
confinement are possible consequences of this.
\end{abstract}
\smallskip

\section{INTRODUCTION}

The idea that among the fundamental laws of physics, we have scale
invariance (s.i.) as one of the fundamental principles appears as
a attractive possibility. In its most naive realizations, such a
symmetry is not a viable symmetry, however, since nature seems to
have chosen some typical scales, so s.i. has to be broken somehow.

While one may look at quantum effects for the origin of scale
symmetry breaking, there is also a way to achieve this at the
classical level in certain models. In the first scale invariant
model of this kind\cite{Guendel1}, a metric, a dilaton field and a
"measure field" $\Phi$ are introduced. The unusual modified
measure\cite{Guendel2} is an object that has the same
transformation properties under general coordinate transformations
as $\sqrt { - g} d^4 x\left( {g = \det \left( {g_{\mu \nu } }
\right)} \right)$. $\Phi$ is a density built out of degrees of
freedom independent of the metric. For example, given a four index
field strength independent of the metric. for example, given a
four index field strength (in four dimensions) $F_{\mu \nu \alpha
\beta }  = \partial _{[\mu } A_{\nu \alpha \beta ]}$, the measure
field $\Phi$ is defined as $\Phi  = \varepsilon ^{\mu \nu \alpha
\beta } F_{\mu \nu \alpha \beta }$. One may consider the field
strength to be composed of elementary scalars\cite{Guendel3}. Four
in the case of a four index field strength, as it was done in
Refs.\cite{Guendel1,Guendel2}.

The scale symmetry breaking in Ref.\cite{Guendel1} comes when
considering the integration of the equation of motion of the four
index field strength, which introduces an arbitrary scale in the
equations of motion. The modified measure idea has been applied to
dynamical generation of string and brane tensions\cite{Guendel4},
to cosmological\cite{Guendel5} and to the fermion family
problem\cite{Guendel6} and to the cosmic coincidences in
cosmology\cite{Guendel7}.

\section{SPONTANEOUS GENERATION OF CONFINING BEHAVIOR}

Let us now turn our attention to gauge theories and consider first
the standard pure Yang-Mills action
\begin{equation}
S = \int {d^4 } x\left( { - \frac{1}{4}F_{\mu \nu }^a F^{a\mu \nu
} } \right), \label{ss10}
\end{equation}
where $ F_{\mu \nu }^a  = \partial _\mu  A_\nu ^a  - \partial _\nu
A_\mu ^a  + gf^{abc} A_\mu ^b A_\nu ^c $. This theory is invariant
under the scale symmetry $A_\mu ^a \left( x \right) \to A_\mu ^a
\left( x \right)^ {\prime } = \lambda A_\mu ^a \left( {\lambda x}
\right)$.

Let us rewrite (\ref{ss10}) with the help of an auxiliary field
$\omega$
\begin{equation}
S = \int {d^4 } x\left( {-\frac{1}{4}\omega ^2 +\frac{1}{2}\omega
\sqrt {-F_{\mu \nu }^a F^{a\mu \nu } } } \right), \label{ss13}
\end{equation}
upon solving the equation of motion obtained from the variation of
$\omega$ and then replacing into the action (\ref{ss13}) we get
back (\ref{ss10}).

Let us consider now\cite{Guendelnu} the replacement $\omega  \to
\varepsilon ^{\mu \nu \alpha \beta } \partial _{[\mu } A_{\nu
\alpha \beta ]}$ and consider now the equation of motion obtained
from the variation of $A_{\nu\alpha\beta}$, which is
\begin{equation}
\varepsilon ^{\gamma \delta \alpha \beta } \partial _\beta  \left(
{\omega  - \sqrt {-F_{\mu \nu }^a F^{a\mu \nu } } } \right) = 0,
\label{ss14}
\end{equation}
which is solved by
\begin{equation}
\omega  = \sqrt {-F_{\mu \nu }^a F^{a\mu \nu } }  + M,
\label{ss15}
\end{equation}
$M$ being once again a space-time constant which produces s.s.b.
of s.i. and in this case it is furthermore associated with the
spontaneous generation of confining behavior. Other examples of
mass generation were studied also in Ref.\cite{Guendelnu} with
similar technique. Indeed the equations of motion obtained from
(\ref{ss13}) in the case $\omega$ is replaced by $\varepsilon
^{\mu \nu \alpha \beta }
\partial _{[\mu } A_{\nu \alpha \beta ]}$, have the form
\begin{equation}
\nabla _\mu  \left( {\left( {\sqrt {-F_{\alpha \beta }^b
F^{b\alpha \beta } }  + M} \right)\frac{{F^{a\mu \nu } }}{{\sqrt
{-F_{\alpha \beta }^b F^{b\alpha \beta } } }}} \right) = 0 .
\label{ss16}
\end{equation}

Now study the equation (\ref{ss16}) for the case of a Abelian
theory and for a spherically symmetric electric field
$F_{0i}=-E_i$ and $F_{ij}=0$, where ${\bf E}=E(r)\hat {\bf r}$.
Then (\ref{ss16}) gives
\begin{equation}
\nabla  \cdot \left( {{\bf E} + \frac{M}{{\sqrt 2 }}\hat {\bf r}}
\right) = 0, \label{ss17}
\end{equation}
which is solved by\cite{GuenGae}
\begin{equation}
{\bf E} =  - \frac{M}{{\sqrt 2 }}\hat {\bf r} + \frac{q}{{r^2
}}\hat{\bf r}. \label{ss18}
\end{equation}
The scalar potential $V$ that gives rise to such electric field is
\begin{equation}
V  =  - \frac{M}{{\sqrt 2 }}r + \frac{q}{r}, \label{ss19}
\end{equation}
which is indeed resembles very much the Cornell confining
potential. Notice that so far (\ref{ss19}) refers to the field of
one charge and not yet to the interaction energy between two
charges. We will see that such interaction energy also has the
Cornell form, even at the quantum level. Since Abelian solutions
are solutions of the non-Abelian theory, these solutions are also
relevant for the non-Abelian generalization.

Before approaching the quantum theory (which will be treated in
some approximations) we want to define effective actions that give
the equations of motion (\ref{ss16}). Indeed one can easily see
that \cite{GuenGae}
\begin{equation}
L_{eff}  = - \frac{1}{4}F_{\mu \nu } F^{\mu \nu }  -
\frac{M}{4}\sqrt { - F_{\mu \nu } F^{\mu \nu }}, \label{ss20}
\end{equation}
reproduces Eqs. (\ref{ss16}).

Since the full treatment of the quantum theory is rather
difficult, instead of using (\ref{ss20}) we restrict ourselves to
a "truncated" phase space model where we consider spherical
coordinates $(r,\theta,\varphi)$ in addition to time, but where we
set $F_{ij}=0=F_{0\varphi}=F_{0\theta}$ and consider only $(t,r)$
dependence of $F_{0r}$. Then instead of (\ref{ss20}), we consider
\begin{equation}
S = 4\pi \int {dr} r^2  L_{eff}, \label{ss21}
\end{equation}
where
\begin{equation}
L_{eff}  = \frac{1}{2}\left( {F_{0r} } \right)^2  - \frac{{M\sqrt
2 }}{4}F_{0r}. \label{ss22}
\end{equation}
Similar kind of "reduced phase space" which take into account only
the spherical degrees of freedom have been used elsewhere in other
examples, see for example Ref.\cite{Benguria}.

\section{INTERACTION ENERGY}

As already mentioned, our aim now is to calculate the interaction
energy between external probe sources in the model (\ref{ss21}).
To do this, we will compute the expectation value of the energy
operator $H$ in the physical state $\left| \Phi  \right\rangle$,
which we will denote by $\left\langle H \right\rangle _\Phi$. The
starting point is the two-dimensional space-time Lagrangian
(\ref{ss22}):
\begin{equation}
L= 4\pi r^2 \left\{ { - \frac{1}{4}F_{\mu \nu } F^{\mu \nu } -
\frac{{M\sqrt 2 }}{8}\varepsilon _{\mu \nu } F^{\mu \nu } }
\right\} - A_0 J^0, \label{ss23}
\end{equation}
where $J^0$ is the external current. A notation remark, in
(\ref{ss23}), $\mu,\nu=0,1$, also, $ x^1  \equiv r \equiv x$ and
$\varepsilon^{01}=1$.

We now proceed to obtain the Hamiltonian. For this we restrict our
attention to the Hamiltonian framework of this theory. The
canonical momenta read $\Pi ^\mu   =  - 4\pi x^2 \left( {F^{0\mu }
+ \frac{{M\sqrt 2 }}{8}\varepsilon ^{0\mu } } \right)$, which
results in the usual primary constraint $\Pi^0=0$, and $\Pi ^i = -
4\pi x^2 \left( {F^{0i}  + \frac{{M\sqrt 2 }}{8}\varepsilon ^{0i}
} \right)$. The canonical Hamiltonian following from the above
Lagrangian is:
\begin{equation}
H_C  = \int {dx} \left( {\Pi _1 \partial ^1 A^0  - \frac{1}{{8\pi
x^2 }}\Pi _1 \Pi ^1  - \frac{{M\sqrt 2 }}{4}\varepsilon ^{01} \Pi
_1  + A_0 J^0 } \right). \label{ss24}
\end{equation}
The consistency condition ${\dot \Pi _0}=0$ leads to the secondary
constraint $\Gamma _1 \left( x \right) \equiv \partial _1 \Pi ^1 -
J^0=0$. It is straightforward to check that there are no further
constraints in the theory, and that the above constraints are
first class. The extended Hamiltonian that generates translations
in time then reads $H = H_C  + \int d x \left( {c_0 (x)\Pi_0 (x) +
c_1 (x)\Gamma _1 (x)} \right)$, where $c_0(x)$ and $c_1(x)$ are
the Lagrange multipliers. Moreover, it follows from this
Hamiltonian that $ \dot{A}_0 \left( x \right) = \left[ {A_0 \left(
x \right),H} \right] = c_0 \left( x \right)$, which is an
arbitrary function. Since $\Pi_0 = 0$, neither $A^0$ nor $\Pi^0$
are of interest in describing the system and may be discarded from
the theory. The Hamiltonian then takes the form
\begin{equation}
H = \int {dx} \left( { - \frac{1}{{8\pi x^2 }}\Pi _1 \Pi ^1  -
\frac{{M\sqrt 2 }}{4}\varepsilon ^{01} \Pi _1  + c^ \prime  \left(
{\partial _1 \Pi ^1  - J^0 } \right)} \right), \label{ss25}
\end{equation}
where $c^ \prime  \left( x \right) = c_1 \left( x \right) - A_0
\left( x \right)$.

According to the usual procedure we introduce a supplementary
condition on the vector potential such that the full set of
constraints becomes second class. A convenient choice is found to
be \cite{Gaete1,Gaete2,Gaete3,Gaete4}
\begin{equation}
\Gamma _2 \left( x \right) \equiv \int\limits_{C_{\xi x} } {dz^\nu
} A_\nu \left( z \right) \equiv \int\limits_0^1 {d\lambda x^1 }
A_1 \left( {\lambda x} \right) = 0, \label{ss26}
\end{equation}
where  $\lambda$ $(0\leq \lambda\leq1)$ is the parameter
describing the spacelike straight path $ x^1  = \xi ^1  + \lambda
\left( {x - \xi } \right)^1 $, and $ \xi $ is a fixed point
(reference point). There is no essential loss of generality if we
restrict our considerations to $ \xi ^1=0 $. In this case, the
only nontrivial Dirac bracket is
\begin{equation}
\left\{ {A_1 \left( x \right),\Pi ^1 \left( y \right)} \right\}^ *
= \delta ^{\left( 1 \right)} \left( {x - y} \right) -
\partial _1^x \int\limits_0^1 {d\lambda x^1 } \delta ^{\left( 1
\right)} \left( {\lambda x - y} \right). \label{ss27}
\end{equation}

We are now equipped to compute the interaction energy between
pointlike sources in the model (\ref{ss25}), where a fermion is
localized at the origin $ {\bf 0}$ and an antifermion at $ {\bf
y}$. As we have already mentioned, we will calculate the
expectation value of the energy operator $H$ in the physical state
$ |\Phi\rangle$. From our above discussion, we see that
$\left\langle H \right\rangle _\Phi$ reads
\begin{equation}
\left\langle H \right\rangle _\Phi   = \left\langle \Phi
\right|\int {dx} \left( { - \frac{1}{{8\pi x^2 }}\Pi _1 \Pi ^1  -
\frac{{M\sqrt 2 }}{4}\varepsilon ^{01} \Pi _1 } \right)\left| \Phi
\right\rangle . \label{ss28}
\end{equation}
Next, as remarked by Dirac\cite{Dirac}, the physical state can be
written as
\begin{equation}
\left| \Phi  \right\rangle  \equiv \left| {\overline \Psi  \left(
\bf y \right)\Psi \left( \bf 0 \right)} \right\rangle  = \overline
\psi \left( \bf y \right)\exp \left( {ie\int\limits_{\bf 0}^{\bf
y} {dz^i } A_i \left( z \right)} \right)\psi \left(\bf 0
\right)\left| 0 \right\rangle, \label{ss29}
\end{equation}
where $\left| 0 \right\rangle$ is the physical vacuum state. As we
have already indicated, the line integral appearing in the above
expression is along a spacelike path starting at $\bf 0$ and
ending $\bf y$, on a fixed time slice.

Taking into account the above Hamiltonian structure, we observe
that
\begin{equation}
\Pi _1 \left( x \right)\left| {\overline \Psi  \left( y
\right)\Psi \left( 0 \right)} \right\rangle  = \overline \Psi
\left( y \right)\Psi \left( 0 \right)\Pi _1 \left( x \right)\left|
0 \right\rangle  - e\int_0^y {dz_1 } \delta ^{\left( 1 \right)}
\left( {z_1  - x} \right)\left| \Phi  \right\rangle. \label{ss30}
\end{equation}
Inserting this back into (\ref{ss28}), we get
\begin {equation}
\left\langle H \right\rangle _\Phi   = \left\langle H
\right\rangle _0  + \frac{{e^2 }}{{8\pi }}\int {dx} \frac{1}{{x^2
}}\left( {\int_0^y {dz_1 \delta ^{\left( 1 \right)} } \left( {z_1
- x} \right)} \right)^2 + \frac{{M\sqrt 2 e}}{4}\int {dx} \left(
{\int_0^y {dz_1 } \delta ^{\left( 1 \right)} \left( {z_1  - x}
\right)} \right), \label{ss31}
\end {equation}
where $\left\langle H \right\rangle _0  = \left\langle 0
\right|H\left| 0 \right\rangle$. We further note that
\begin{equation}
\frac{{e^2 }}{2}\int {dx} \left( {\int_0^y {dz\delta ^{\left( 1
\right)} \left( {z_1 - x} \right)} } \right)^2  = \frac{{e^2
}}{2}L , \label{pato}
\end{equation}
with $|y|\equiv L$. Inserting this into Eq.(\ref{ss31}), the
interaction energy in the presence of the static charges will be
given by
\begin{equation}
V =  - \frac{{e^2 }}{{8\pi }}\frac{1}{L} + \frac{{M\sqrt 2
e}}{4}L, \label{ss32}
\end{equation}
which has the Cornell form. In this way the static interaction
between fermions arises only because of the requirement that the
$\left| {\overline \Psi \Psi } \right\rangle$ states be gauge
invariant.

\end{document}